\documentclass[twocolumn,preprintnumbers,amsmath,amsfonts,amssymb,notitlepage,showpacs,pra]{revtex4-1}
\usepackage{geometry}
\usepackage{color}
\date{\today}
\DeclareMathAlphabet{\mathpzc}{OT1}{pzc}{m}{it}
\usepackage[utf8x]{inputenc}
\usepackage{amsmath}
\usepackage{amsfonts}
\usepackage{amssymb}
\usepackage{graphicx}
\usepackage{nicefrac}
\usepackage{amsbsy}
\usepackage{float}
\usepackage{epstopdf}
\usepackage{dsfont}
\usepackage{siunitx}
\usepackage{lipsum}
\usepackage{graphicx}
\usepackage{tikz}
\newcommand*\circled[1]{\tikz[baseline=(char.base)]{
            \node[shape=circle,draw,inner sep=0.5pt] (char) {#1};}}

\def \be{\begin{equation}}
\def \ee{\end{equation}}
\def \ba{\begin{array}}
\def \ea{\end{array}}
\def \bea{\begin{eqnarray}}
\def \eea{\end{eqnarray}}

\begin{document}
\title{
Driven-dissipative control of cold atoms in tilted optical lattices
}
\author{Vaibhav Sharma}
\email{vs492@cornell.edu}
\author{Erich J Mueller}
\email{em256@cornell.edu}
\affiliation{Laboratory of Atomic and Solid State Physics, Cornell University, Ithaca, New York}
\begin{abstract}
    We present a sequence of 
    driven-dissipative protocols for controlling cold atoms in tilted optical lattices.  These
    experimentally accessible examples
    are templates that demonstrate how dissipation can be used to manipulate quantum many-body systems.
    We consider bosonic atoms trapped in a tilted optical lattice, immersed in a superfluid bath, and excited by coherent Raman lasers. 
    With these ingredients, we are able to controllably transport atoms  in the lattice and produce self-healing quantum states:  a Mott insulator and the topologically ordered spin-1 AKLT state.
\end{abstract}

\maketitle

\section{Introduction}

While dissipation is often viewed as a hindrance, it can also be a tool for 
manipulating quantum states.
Here we provide a series of cold atom examples of using dissipation to prepare quantum states, and
control their behavior.
%
%
This approach complements other methods of state preparation, and avoids some of the challenges those techniques face. By giving specific protocols for dissipative state preparation and control, we are able to elucidate the underlying principles and directly confront the advantages and disadvantages of this approach.

In cold atom experiments, dissipation is provided by several mechanisms: off-resonant light scattering, three-body atom loss, collisions with background atoms, and even the conversion of coherent excitations into incoherent ones through elastic collisions of atoms in the sample.  Traditional cooling schemes have long relied on these processes for equilibration~\cite{coolmethod}, but the idea of engineering dissipation to target specific many-body quantum states is more novel~\cite{engnrdissipation}.  In the few-particle limit, there are more established examples, such as optical pumping.  The driven-dissipative approach to state preparation can be viewed as a many-body analog of optical pumping.

This method 
offers a practical alternative to other  techniques of state preparation, and overcomes some of their difficulties. For instance, adiabatic state preparation requires the process to be much slower than the many-body gap~\cite{adiabatictheorem}. This poses particular challenges if one needs to traverse a critical region, where the gap vanishes. Another common state preparation technique is Hamiltonian engineering where the desired state is the ground state of some fixed Hamiltonian which can be reached by cooling.  
Obstacles to Hamiltonian engineering include:
(1) The required temperatures may be unachievably low, and (2) the equilibration
rates may become small as one approaches the state of interest.
An example is the Fractional Quantum Hall state, where particles avoid one-another, and hence the elastic collision rate is small.  In all of these techniques, the key questions are: (1) Can you produce the state of interest, and (2) How long does it take?  We answer both  these questions  for our examples.

Dissipative state preparation has the potential to be fast -- the timescale is experimentally controllable and the kinetic bottlenecks can be explicitly removed by properly engineering the environment.  The critical slowing down that plagues adiabatic methods is completely  avoided: one is  far from equilibrium throughout the time evolution.  Most importantly, the final state is self-correcting.  If one leaves on the dissipation (perhaps with some reduced amplitude) any perturbations can be healed.  This has connections to autonomous error correcting codes: the dissipatively stabilized quantum state is a protected resource for manipulating quantum information.


Controlled driven-dissipative state preparation is an active area of research both theoretically and experimentally.
For example, in superconducting qubits, dissipation has been used to create a stable Mott insulator of photons~\cite{mottphoton} and a long lived two qubit Bell state~\cite{bell}. Reservoir engineering has been used in trapped ion systems  to create a four qubit GHZ state~\cite{GHZ}. Ultracold atoms can be promising candidates for dissipative state preparation. The Mott-superfluid transition in a driven-dissipative Bose-Hubbard system has been realized experimentally~\cite{bosehubbard}. Some theoretical proposals include preparation of number and phase squeezed bosonic states~\cite{squeeze} and dissipatively prepared topological superconductors~\cite{toposupercond}. 

In this work, we analyze a broadly applicable approach to driven-dissipative control in cold atoms, which both extends these examples and provides an experimentally accessible framework for exploring the general principles.  Our setup is schematically shown in Fig.~\ref{coherentdrive}.
%
We consider ${}^7$Li atoms  in a ``tilted" 
one dimensional optical lattice, modelled by the sum of a linear and sinusoidal potential. Transitions are driven by two-photon Raman processes, which use an electronically excited state as an intermediary in changing the spatial mode of an atom.   Dissipation is provided by coupling to a superfluid bath of $^{23}$Na atoms which are not trapped by the lattice. 
Lithium atoms can decay from excited vibrational states to lower ones by emitting Bogoliubov excitations in the superfluid bath.
We note that all the different parts of our proposed process 
have been experimentally demonstrated in other works \cite{tilt,maggradient,WScoherent,DeMarcobath,NaLi}.  Similarly, these ingredients have been studied in a number of theory works \cite{Zollerbath,specdeplatt}.
We explicitly calculate all the relevant energy and time scales and quantitatively show how the drive and dissipation can produce our target states. 

We give three examples to demonstrate the capabilities of our driven-dissipative scheme. First, we show a way to control atom transport along a  one dimensional tilted optical lattice potential, both up and down the lattice. This would be relatively simple to implement and an excellent target for initial experiments. Additionally, it provides a way to fully control the speed of atom transport in a lattice. Second, we show how it can be further modified  to drive the system into a strongly-correlated Mott insulator state in a tilted lattice with a mechanism to self-heal any holes. Finally, we propose how the iconic AKLT (Affleck-Kennedy-Lieb-Tasaki) state can be achieved with this technique.

The AKLT state occurs in a spin-1 chain.  It has unique properties, including symmetry protected topological order and valence bond order. It has emergent spin-1/2 edge modes and a gapped ground state. It is also a prototypical example of a matrix product state. The AKLT state has never been created experimentally in cold atom systems. One of our central results is a protocol to create it in a driven-dissipative ultra cold atomic system.

Our paper is organized as follows. In Section II we introduce the physical system and 
derive the effective model which we will use to describe it.
This includes showing how the driving and dissipation processes are engineered. In Section III, we describe our protocols and
calculate state preparation time,
including  the scaling with system size.


\section{Model}

\subsection{System and Effective Model}

As already introduced, our basic setup is shown in Fig.~\ref{coherentdrive}.
For each of our examples, we consider two populations of bosons, referred to as the lattice and the bath atoms. The lattice atoms are constrained to move in one dimension (1D).  They experience a ``tilted lattice" potential along that direction, consisting of the superposition of  sinusoidal and  linear potentials.  The bath atoms form a 3D cloud.  There are also a series of control lasers that are used to drive Raman transitions.  Each of these components are discussed in detail below.

In its simplest incarnation, the resulting effective model has the structure of a 
 1D chain of sites with a linear potential gradient.  Each site $j$ contains two states:  The ground state $|g\rangle_j$ and a vibrationally excited state $|e\rangle_j$, with an energy gap, $\hbar\omega_0$ between them.
 \begin{equation}\label{h0}
\hat H_{0} = \sum_{j} -j\Delta|g\rangle_j \langle g|_j + (-j\Delta + \hbar\omega_0)|e\rangle_j \langle e|_j.
\end{equation}
 Additionally, there is an onsite interaction Hamiltonian, 
 \begin{equation}\label{hint}
\hat H_{\rm int} = U_{gg} |gg\rangle\langle gg|+
U_{ge}|ge\rangle\langle ge|
+ U_{ee} |ee\rangle\langle ee|,
\end{equation}
and coherent drives, 
\begin{equation}\label{hcoh}
H_{\rm coh} = \Omega^{\prime}e^{-i\omega t}|g\rangle_j \langle e|_{j+1} +h.c.
\end{equation}
Our decay terms are modelled by on-site jump operators of the form,
\begin{equation}\label{qjump}
\hat{L}_j = \sqrt{\Gamma}|g\rangle_j\langle e|_{j}.
\end{equation}

Subsections~\ref{1dlat} through \ref{limitations} derive this effective model, starting with a microscopic description of the trapped gas.

\subsection{1D Lattice}\label{1dlat}

\begin{figure}
\includegraphics[width=\columnwidth]{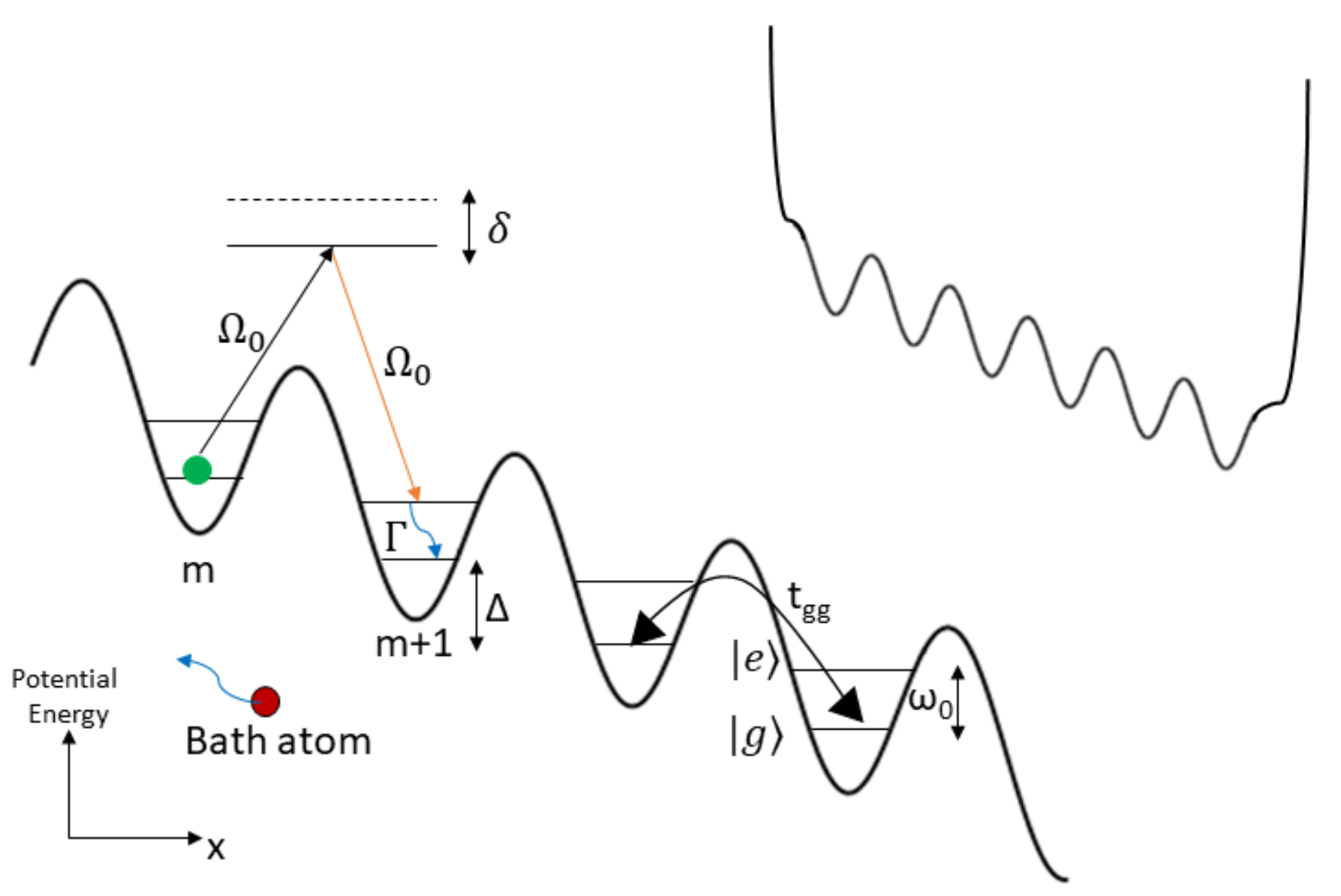}
\caption{(color online) Spatial dependence of potential energy in a one dimensional tilted optical lattice.   Each site has two eigenstates, labelled $|g\rangle$ and $|e\rangle$ with an energy gap $\omega_0$. Energy offset between neighboring sites is $\Delta$ and the tunneling within the lowest band is $t_{gg}$.  As depicted by black and orange arrows, labeled by $\Omega_0$, coherent Raman lasers drive transitions  
between the ground band of site $m$ and the excited band of site $m+1$.
The Raman transition is detuned by $\delta$ from an intermediate electronic excited state.
The lattice is immersed in a 3D superfluid bath (bath atom shown in red). The blue curvy arrow shows spontaneous decay from excited band to ground band through bath interactions with a rate, $\Gamma$.  
Upper right: A finite length tilted lattice with the additional box potential walls at the edges.}
\label{coherentdrive}
\end{figure}

The system atoms are trapped in 
 a one dimensional tilted optical lattice. The lattice is generated by interfering two counter-propagating laser beams.   The tilt can be generated by using the AC stark shift from a large waist laser which is incident on the system from a direction perpendicular to the lattice~\cite{tilt}. The resulting potential is shown in Fig.~\ref{coherentdrive}.

When the energy difference between neighboring sites, $\Delta$, is large compared to the tunneling amplitude, $t$, then the eigenstates of the tilted lattice become strongly localized to individual sites.  If we limit ourselves to nearest neighbor hopping, the exact Wannier-Stark eigenstates are \cite{Wstark}
\begin{equation}\label{WSfns}
\psi_{m,\alpha} = \sum_{l} J_{l-m}(\frac{t_\alpha}{\Delta}) \phi_l^\alpha,
\end{equation}

Here, $\alpha$ is the band index, $l$, $m$ are site indices, $\phi_l^\alpha$ is the site localized Wannier function for the $\alpha$'th band, and $t_\alpha$ is the hopping matrix element for that band. $J_n(x)$ is the $n$'th Bessel function: for small arguments, $J_n(x)\sim x^{|n|}$, and we see that the wavefunction falls off exponentially.
For a sufficiently tight lattice, one can approximate the Wannier states, $\phi_m^\alpha$, as harmonic oscillator eigenstates, with energy
\begin{equation}\label{eband}
E_{m,\alpha} = -m\hbar\Delta + (\alpha + 1/2)\hbar\omega_0,
\end{equation}
where $\omega_0$ is the small oscillation frequency.
Throughout we will only consider two bands, labeled by $\alpha=g,e$ -- and regardless of the accuracy of the harmonic approximation, we can define $\hbar \omega_0=E_{m,e}-E_{m,g}$.

For two particles on a site,
the on-shell components of the on-site interaction are,
\begin{equation}
\hat H_{\rm int} = U_{gg} |gg\rangle\langle gg|+
U_{ge}|ge\rangle\langle ge|
+ U_{ee} |ee\rangle\langle ee|
\end{equation}
where $|\alpha\beta\rangle$ is the state with particles in bands $\alpha$ and $\beta$, and the site index has been suppressed.
The $U$'s scale as $a_s/d_\perp^2$, where $a_s$ is the scattering length, and $d_\perp$ is the transverse size of the Wannier state.
If one approximates the Wannier states as harmonic oscillator eigenstates, then $U_{gg}$ can be calculated as, 
\begin{equation}
    U_{gg} = \frac{8\pi\hbar^2 a_s}{m d_\perp^2}\int_{-\infty}^{\infty} dx\, |w_{g}(x)|^4
\end{equation}
Here, $w_g$ is the ground state harmonic oscillator wavefunction. Analogous expressions give the interaction energies involving higher bands, and one finds  $U_{ge} = \frac{1}{2}U_{gg}$ and $U_{ee} = \frac{3}{4}U_{gg}$.  

We model the lattice potential as
$V(r) = V_0 E_{\rm rec} \sin^2(2\pi x/d)$,
where $V_0$ is the dimensionless lattice strength, $d$ is the lattice beam wavelength, and $E_{\rm rec} = \frac{2\hbar^2 \pi^2}{m d^2}$ is the recoil energy.  This potential has a lattice spacing of $d/2$. 

The tunneling amplitude, $t_{gg}$ in the lowest motional band in the harmonic approximation can be written as, 
\begin{eqnarray}
    t_{gg} &=&
    \int dx\, w_g^*(x+d/2) \hat H w_g(x)
    \\
    &\sim& E_{\rm rec} \sqrt{V_0}e^{-\frac{\pi^2}{4}\sqrt{V_0}}
\end{eqnarray}

The tunneling in the higher band can be similarly found to be, $t_{ee}\sim \frac{3 t_{gg}}{\sqrt{2}} (1+\frac{\pi^2}{4}\sqrt{V_0})$.
The bandgap between the ground and excited motional band is, 
\begin{equation}
   \hbar \omega_0 \sim 2E_{\rm rec}\sqrt{V_0}.
\end{equation}
As a rule of thumb, these approximations work well when $V_0 \gtrsim 9$.
For $d=1064$nm, a typical laser wavelength, one then has
a tunneling amplitude on the order of tens to hundreds of Hz, while the band gap between the ground and first band is on the order of tens of kHz. Typical background scattering lengths are of the order of a few nm for ${}^7$Li atoms~\cite{Liswave}. These scattering lengths can be easily tuned via Feshbach resonances, with the caveat that one may find enhanced inelastic processes if they are made too large. In our system, we envision that the lattice atoms are tightly confined in the transverse directions with $d_\perp \ll d$ and thus the interaction energy is also on the order of tens of kHz.

Such one dimensional tilted optical lattices have been realized in experiments either using the AC stark shift gradient from a laser~\cite{tilt} or with a magnetic field gradient~\cite{maggradient}.

\subsection{Coherent Drive}\label{drivesection}

Transitions are driven by a two-photon
coherent drive:  the lasers are tuned so that absorbing a photon from one beam, and emitting it into the second is resonant with a band-changing hopping event.  For example, as illustrated in Fig.~\ref{coherentdrive}, this Raman process could drive
the transition from the ground band on one site to the excited band on a neighboring site, in which case 
$\omega_1-\omega_2=\omega_0-\Delta$.  Here $\omega_1=c |k_1|$ and $\omega_2=c |k_2|$ are the frequencies of the lasers.
An incoherent scattering event will later return the atom to the ground band.

The Raman laser frequencies can be adjusted to bring other possible transitions like hopping between ground bands of neighboring sites into resonance.

In our concrete scenario, 
one laser beam drives the atom in the lowest band in site $m$ to a virtual level, corresponding to an electronically excited state. This single photon process is detuned by frequency $\delta$, and has transition rate  $\Omega_0$. The second laser drives the transition from this virtual level to the motionally excited state of site $m+1$.


If $\delta \gg \Omega_0$, the effective rate of this two photon process can be calculated by adiabatically eliminating the higher electronic level and is given by:
\begin{eqnarray}
\label{ramanrate}
    \Omega^{\prime} &=& \frac{\Omega_0^2}{\delta}
    \left|\int dx\,
    \psi_{m,g}^*(x)
    e^{i(k_{1}-k_{2})x}\psi_{m+1,e}(x) 
    \right|^2\\
    \label{ramanratefinal}
    &\sim& \frac{\Omega_0^2}{\delta} \left(1+ \frac{\pi^2\sqrt{V_0}}{2} \right) e^{-\frac{\pi^2\sqrt{V_0}}{2}}.
\end{eqnarray}
The latter result is derived in Appendix~\ref{ramanapp}, in the deep lattice limit, with the angle between the Raman beams chosen to optimize the transition rate.  

To make unwanted transitions highly off-resonant, we want to be in the regime where $\Omega^{\prime} \ll \Delta , \omega_0$ (which as previously explained are on the order of $10^4$Hz). 
 

These conditions would ensure that our drive induces coherent Rabi oscillations of atom transfer between adjacent sites. A similar coherent atom transfer process in a Wannier-Stark ladder has been experimentally demonstrated by Beaufils et al.~\cite{WScoherent}.  They were able to achieve Raman transition rates larger than needed for our proposal.

\subsection{Dissipation Medium}

Dissipation is provided by scattering from a particle bath of superfluid bosons that do not feel the optical lattice
~\cite{Zollerbath}. An atom in the higher band can transition to the ground band through spontaneous emission of a Bogoliubov  excitation in the superfluid bath. This spontaneous decay process is shown schematically in Fig.~\ref{coherentdrive}.

The temperature of the superfluid bath is much smaller than $\omega_0$, so the bath of Bogoliubov particles can effectively be taken as $T=0$:  These excitations can be created, but are never absorbed.



Fermi's golden rule can be used to calculate the resulting decay rate from the motionally excited state to the ground state.  
Following the argument of Griessner {\em et al.}~\cite{Zollerbath}, we find the decay rate in our system to be,
\begin{equation}\label{decayrate}
\begin{split}
    \Gamma &= 8\pi\rho_b a_{ab}^2 \sqrt{E_{rec.}}\frac{V_0^{1/4}}{m_a^{1/2}} e^{-\frac{4\pi^2 m_r \sqrt{V_0}d_\perp^2}{d^2}}f(m_r)\\
    f(m_r) &= \frac{(1+m_r)^2}{m_r} (\sqrt{\pi}\mathrm{erf}(\sqrt{m_r}) - 2\sqrt{m_r}e^{-m_r}) 
\end{split}
\end{equation}
Here, $m_b$ and $m_a$ are the masses of the bath atoms and lattice atoms respectively, and $m_r = \frac{m_b}{m_a}$ is their ratio.
The interactions between the lattice and bath atoms are parametrized by the s-wave scattering length, $a_{ab}$. 
The number density of bath atoms is
$\rho_b$, and $\mathrm{erf}(x)$ is the error function. 

The rate can be tuned by changing the superfluid bath density, scattering length and atomic mass ratios. The first of these is experimentally the most accessible -- but Feshbach resonances can be used to control $a_{ab}$.

The rate strongly depends on the mass ratio, $m_r$. For $m_r \ll 1$, $f(m_r)$ scales as $\sqrt{m_r}$ while for $m_r \gg 1$, the rate exponentially decays as $e^{-\frac{4\pi^2 m_r \sqrt{V_0}d_\perp^2}{d^2}}$. The rate exponentially decreases at larger mass ratios because the energy transfer is poorer in collisions with higher mass atoms. The optimal mass ratio occurs for $m_r\sim 10$ [for experimentally typical values of $\frac{\sqrt{V_0} d_\perp^2}{d^2}$]. 

For concreteness, we consider the ${}^7$Li atoms in the lattice immersed in a superfluid bath of ${}^{23}$Na atoms. For a typical bath atom density of $10^{13} \;{\rm cm}^{-3}$ and inter-species scattering length on the order of a few nm, the decay rate is on the order of a few kHz. This rate sets the time-scale of the experiment.  Our coherent transfers rely upon
resolving the motional sidebands and the lattice tilts, and hence require $\Gamma \ll \Omega^\prime,\omega_0,\Delta, U$.  

In this setup, the bath atoms do not feel the optical lattice:
this can be arranged due to the different AC polarizabilities of the atomic species. For example, R. Scelle {\em et al.} \cite{NaLi} immerse Lithium atoms in a condensate of Sodium atoms where only the Lithium atoms are trapped in a species-specific optical lattice. A detailed discussion of techniques to create species-specific optical lattices for various alkali atom mixtures is given by LeBlanc and Thywissen in \cite{specdeplatt}.

This decay process can be viewed as a form of sympathetic cooling -- and has been experimentally studied by Chen et al.~\cite{DeMarcobath} in that context.

\subsection{Additional Potentials}\label{endbarrier}
For some of our protocols we also  add  an additional potential, which can be created with an off-resonant laser.  In particular, we wish to be able to create a finite length chain as shown in the upper right corner in Fig.~\ref{coherentdrive}, by adding barriers at the end of the chain.  Such potentials are commonly generated in experiments~\cite{wall}.

\subsection{Limitations of Effective Model}\label{limitations}

An experiment can only be described by the effective model in Eqs.~(\ref{h0}) through (\ref{qjump}) if  $\Delta,\omega_0 \gg \Omega^{\prime},\Gamma$.

The condition $\Delta,\omega_0 \gg \Omega^{\prime}$ is required so that the drive does not produce transitions to unwanted sites.  For example, the drive inevitably produces a matrix element connecting $|g\rangle_j$ and $|g\rangle_{j+1}$.  This transition is off-resonant, though, and the rate is suppressed by a factor of $\Omega^\prime/\omega_0$ relative to the wanted transition.  Similarly, the matrix element connecting $|g\rangle_j$ and $|e\rangle_j$ is suppressed by $\Omega^\prime/\Delta$.



The condition $\Delta,\omega_0 \gg \Gamma$ is required so that the level broadening does not bring any of those same unwanted transitions into resonance.


Note, these conditions puts constraints on the techniques which can be used to introduce the dissipation.  For example, it would be challenging to design the protocols so that spontaneous emission of a photon would provide the dissipation.  The characteristic scales of optical processes are much larger than $\Delta$ and $\omega_0$.  The scales of the Bogoliubov excitations are better-matched.




\section{Proposals}
As already explained, we propose three scenarios:  In section~\ref{elevator}, we describe a novel transport scheme where this driven-dissipative approach controls the motion of  a cold gas.  In section~\ref{Mott}, we show a variant of the technique that can be used to heal defects in a Mott state.  Finally, in section~\ref{AKLT}, we explain how to pump the system into the AKLT state  -- a highly nontrivial example of state engineering.

All of these will be described using variants of the model introduced in Eqs. (\ref{h0}) through (\ref{qjump}).

\subsection{Raman Sideband Elevator}
\label{elevator}

Transport in solid state systems is a driven-dissipative process.  A potential gradient provides energy to the system, while inelastic scattering off of impurities acts as a regulator, controlling the average speed of the electrons.  We propose constructing the analogous process in our atomic system.  As in the solid-state system, the atoms will move with constant velocity.  By changing the intensity and frequency of the Raman lasers, one can control both the direction and speed of motion, which leads us to refer to this as a ``Raman Sideband Elevator."


As depicted in Fig.~\ref{coherentdrive}, the coherent Raman drive  resonantly couples atoms from ground band of one site to the excited band of its nearest neighbor. The atom in the higher band interacts with the dissipative bath and decays to the ground band at that site. This irreversibly transfers the atom one site down the ladder. The process can then repeat itself. 
In steady state, all atoms are moving at a constant speed.
This elevator can transfer atoms in either direction, depending on the frequencies of the Raman lasers: when $\omega_1-\omega_2=\omega_0-\Delta$, the transfer is down-hill, while when $\omega_1-\omega_2=\omega_0+\Delta$, the transfer is uphill. 
The analysis is identical, and throughout this section, we use the notation appropriate for down-hill transport.
%


We will neglect interactions between atoms, and just model the single-particle problem.  The weakly interacting regime is readily reached by either reducing the strength of transverse confinement or using a  Feshbach resonance. 

Given that the single-particle eigenstates are localized on each site, this situation can be modelled via a classical rate equation:  the only relevant variable is $n_i$, the expectation value of the number of particles on site $i$, and
\begin{equation}\label{rate}
    \frac{dn_i}{dt} = \tilde\Gamma n_{i-1} -\tilde\Gamma n_i.
\end{equation}
The rate $\tilde\Gamma$ is found by modeling the three levels involved in the transport of atoms by one site.   Under the regime where the driving rate is greater than the decay rate, that is, $\Omega^\prime \gg \Gamma$, one finds $\tilde\Gamma=\Gamma/2$.  The Raman lasers cause the atom to execute many Rabi oscillations, so it spends half of its time in the unstable state -- which decays with rate $\Gamma$.


These rate equations can be solved exactly to determine the speed at with which the center of mass moves down-hill and how much the cloud spreads over time. Using Eq.~(\ref{rate}), the rate of change of the center of mass position of the atom cloud, $X_{com}$ is given by, 
\begin{equation}
    \frac{d X_{com}}{dt} = \frac{1}{\sum_i n_i}\frac{d(\sum_i i n_i)}{dt} = \tilde\Gamma
\end{equation}

Similarly, the rate of change of the cloud spread, $\sigma^2$ can be derived from Eq.~\ref{rate} to be, \begin{equation}
    \frac{d\sigma^2}{dt} = \frac{1}{\sum_i n_i}\frac{d(\sum_i i^2 n_i - (\sum_i i n_i)^2)}{dt} = \tilde\Gamma
\end{equation}

Under this process, the atom cloud's center of mass moves downhill with a constant speed, controlled by the dissipation rate, $\tilde\Gamma$. At the same time, there is a linear spread of the square of the cloud size with time. Both the drift and the spread are readily measured in experiments.

One can also envision a finite length chain with all the atoms initially confined at the upper end of the ladder, and a potential barrier at the bottom, as described in Sec.~\ref{endbarrier}. The final steady state would be reached when all atoms accumulate in the last site downhill. For a chain of length $L$, the time taken would scale as, $t \sim L/{\tilde\Gamma}$.

\subsection{Mott State}\label{Mott}

An ideal Mott insulating state contains exactly one particle per site.  This is the ground state of the Bose Hubbard model with very strong interactions~\cite{mottstate}.  Finite temperature introduces defects, as do atom loss events.  A variant of our Raman Sideband Elevator can pump the system into an ideal Mott state, and heal defects which are later created.

We consider a finite length chain, with a potential similar to the one in the upper right corner in Fig.~\ref{coherentdrive}. We require that the atom-atom interaction, $U_{ge}\gg \Gamma,\Omega^\prime$.  This large interaction strength can be engineered by tightening the transverse confinement, or using a Feshbach resonance.

\begin{figure}
\includegraphics[width=7cm]{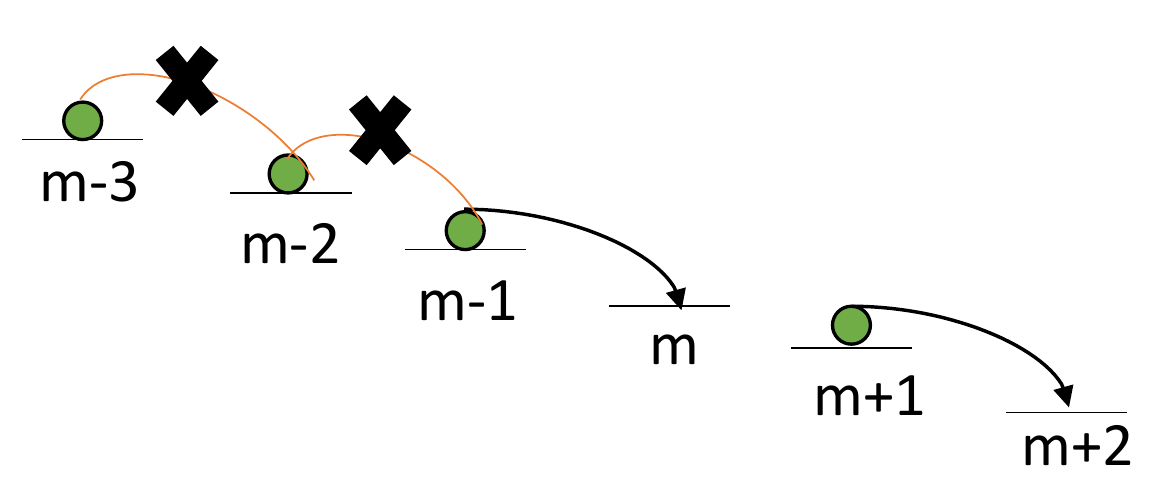}
\caption{A partially filled configuration.
The driven dissipative process leads to incoherent hopping in the down-hill direction.  In the presence of strong interactions, hopping onto a filled site is detuned and therefore forbidden.  This is indicated by the red arrows with  x's through them.
}
\label{mottfig}
\end{figure}

With strong interactions amongst the lattice atoms, the Raman transition between the ground state of site $i$ and the excited state of site $i+1$ is only resonant if there are no atoms on site $i+1$.  Effectively, this means our incoherent hopping only occurs onto an empty site.
 Starting from a sparsely filled chain where the average occupation per site is less than 1, as shown in Fig.~\ref{mottfig}, the driven-dissipative process enables transport of the atoms down the ladder from filled sites to empty sites.  The right-most atom stops when it hits the barrier.  Subsequent atoms stop when they encounter the filled states.  The end result is an idealized Mott state, with one atom per site.  
 If holes later develop, they are rapidly filled, as all uphill particles  shift over by one site.

We can estimate the time needed for a partially filled ladder 
to reach the final Mott state.
The model from Section~\ref{elevator} is simply modified to forbid hopping onto an occupied site. We use a Gillespie algorithm to simulate the resulting stochastic dynamics \cite{gillespie}.

We find that the results are very well approximated by assuming that the probability distributions on different sites are uncorrelated and incoherent hopping is only allowed on empty sites.  Thus the dynamics are efficiently simulated using the 
``mean field'' rate equation
\begin{equation}
    \frac{dn_i}{dt} = \tilde\Gamma n_{i-1}(1-n_i) - \tilde\Gamma n_i(1-n_{i+1})
\end{equation}
%
As before, $n_i$ is the expectation value of the number of particles on site $i$.

For a finite length chain, the infinite time solution of the above equation is the desired Mott state, where all atoms are jammed up at the right hand side. For generic initial conditions, the time to reach this dark state scales linearly with the size of the system.



Double occupancies are highly disruptive here.  Our procedure does not allow them to move, and they act as barriers that prevent the motion of other atoms.  There are several techniques to remove doubly occupied sites~\cite{doublon}.

After the ideal Mott state is formed, an atom loss event will create a hole.  This is healed by the hole hopping to the left.  The hole needs to hop at most $N$ sites, where $N$ is the number of particles.  Thus  the characteristic time for the repair is $\tau\sim N/\tilde\Gamma$.


\subsection{AKLT State}\label{AKLT}

Finally, a variant of this set-up can be used to create the AKLT (Affleck-Kennedy-Lieb-Tasaki) state of the spin-1 chain. As previously explained, the AKLT state is a symmetry protected topologically ordered state having topologically protected edge modes.

The set-up here is slightly more involved than our previous examples, as we need to manipulate the hyperfine spin degrees of freedom.  
We consider the construction demonstrated in experiments at MIT \cite{Ketterletwocomp} where they
build local spin-1 objects by placing two atoms on each site.  Each atom has two accessible hyperfine states, and is effectively a spin-1/2 object.  Because two Bosons in the same site must be in a spin symmetric state, these form a spin-1 composite.  In the experiments, bosonic $^7$Li was used.

\begin{figure}
\includegraphics[width=7cm]{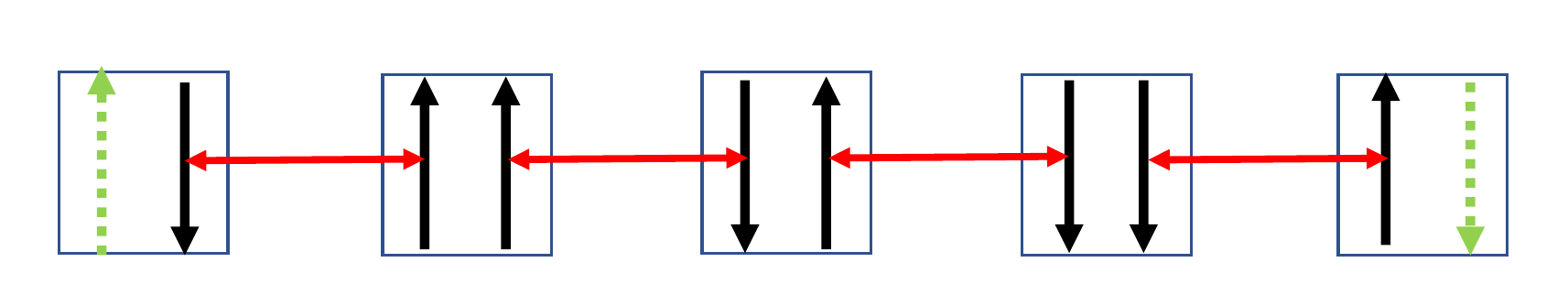}
\caption{AKLT state described in a {\em parton} picture, where the spin-1's are represented by two spin-1/2 particles at each site. Each blue square corresponds to projection of the two spin-1/2's into a triplet state while each red double arrow connecting spin-1/2's on neighboring sites represents a singlet. The green (dashed) spins at the two ends are spin-1/2 edge modes.}
\label{akltfig}
\end{figure}

The physical structure in this experiment parallels the  ``parton" construction often used to understand the AKLT state \cite{aklt}.  In that picture, each spin-1 is broken into two (symmetrized) spin-1/2's as shown in Fig.~\ref{akltfig}.  Each of these spin-1/2's forms a singlet with a parton from a neighboring site.  
The edge modes are understood as the leftover spins which do not have singlet partners.  The partons are usually a mathematical construct, but in the experiment they represent individual atoms.

In the original spin-1 language, the AKLT state
is uniquely (up to the boundary modes) defined by the property that 
if you take any two neighboring spin-1's -- they have zero weight in the spin-2 channel.  In other words, the AKLT state is annihilated by the operators $P^{(S_T=2)}_{i,j}= \frac{1}{2}\mathbf{S_i}\cdot \mathbf{S_j}+ \frac{1}{6}(\mathbf{S_i}\cdot\mathbf{S_j})^2 +\frac{1}{3}$, each of which projects the spins on neighboring sites $i$ and $j$ into spin-2.  The simultaneous null-space of these operators is 4-fold degenerate, corresponding to the two spin-1/2 edge degrees of freedom.  The AKLT state is one of the prototypical examples of a matrix product state. It has also been proposed as a potential platform for measurement based quantum computing~\cite{quantumcomputing}, and this projector construction has analogs with stabilizer codes~\cite{stabilizer}.

Our goal is to engineer a dissipative process which occurs only when two neighboring sites are in the spin-2 sector.  The AKLT state will then be the unique dark state.  Again, uniqueness is only up to the boundary mode configuration.  The strategy will be to have the dissipation 
involve
an intermediate state with four bosons on a single site -- a configuration which can only be reached if the atoms are in the spin-2 sector.

Each site can be in one of the following three triplet states, which are the different $S_z$ projections of spin 1:
\begin{eqnarray}
    |\uparrow \uparrow \rangle = |+\rangle\\
    \frac{1}{\sqrt{2}}|\uparrow \downarrow + \downarrow \uparrow  \rangle = |0\rangle\\
    |\downarrow \downarrow \rangle = |-\rangle
\end{eqnarray}
As seen in~\cite{Ketterletwocomp}, each of the spin-1 states on a site, $|+\rangle$, $|0\rangle$ and $|-\rangle$ have spin-dependent on-site interaction energies, $U_{gg}^{|\sigma\rangle}$, with $\sigma=-,+,0$. These interaction energies depend on the magnetic field.
Amato-Grill {\em et al.}~\cite{Ketterletwocomp} find that there are special values of magnetic fields where the three $S_z$ projections have equally spaced interaction energies, that is, $U_{gg}^{|+\rangle}- U_{gg}^{|0\rangle} = U_{gg}^{|0\rangle}-U_{gg}^{|-\rangle} = \Delta U_{gg}\sim$ kHz.  Under those conditions, the spin dependence of the interactions are equivalent to a magnetic field in the $\hat z$ direction. In our dynamics, the total $S_z$ will be conserved, so this field plays no role.  The spin Hamiltonian is then effectively rotationally invariant.  Note that $\Delta U$ is small compared to $U_{gg}^{|\sigma\rangle}\sim U_{gg}$, which is tens of kHz.  

The same story works in the excited band, but $U_{ee}^{|\sigma\rangle}\neq U_{gg}^{|\sigma\rangle}$.
Within our approximations, $U_{ee}^{|\sigma\rangle}=(3/4) U_{gg}^{|\sigma\rangle}$, which means
$\Delta U_{ee}=(3/4)\Delta U_{gg}$.


\begin{figure}
\includegraphics[width=\columnwidth]{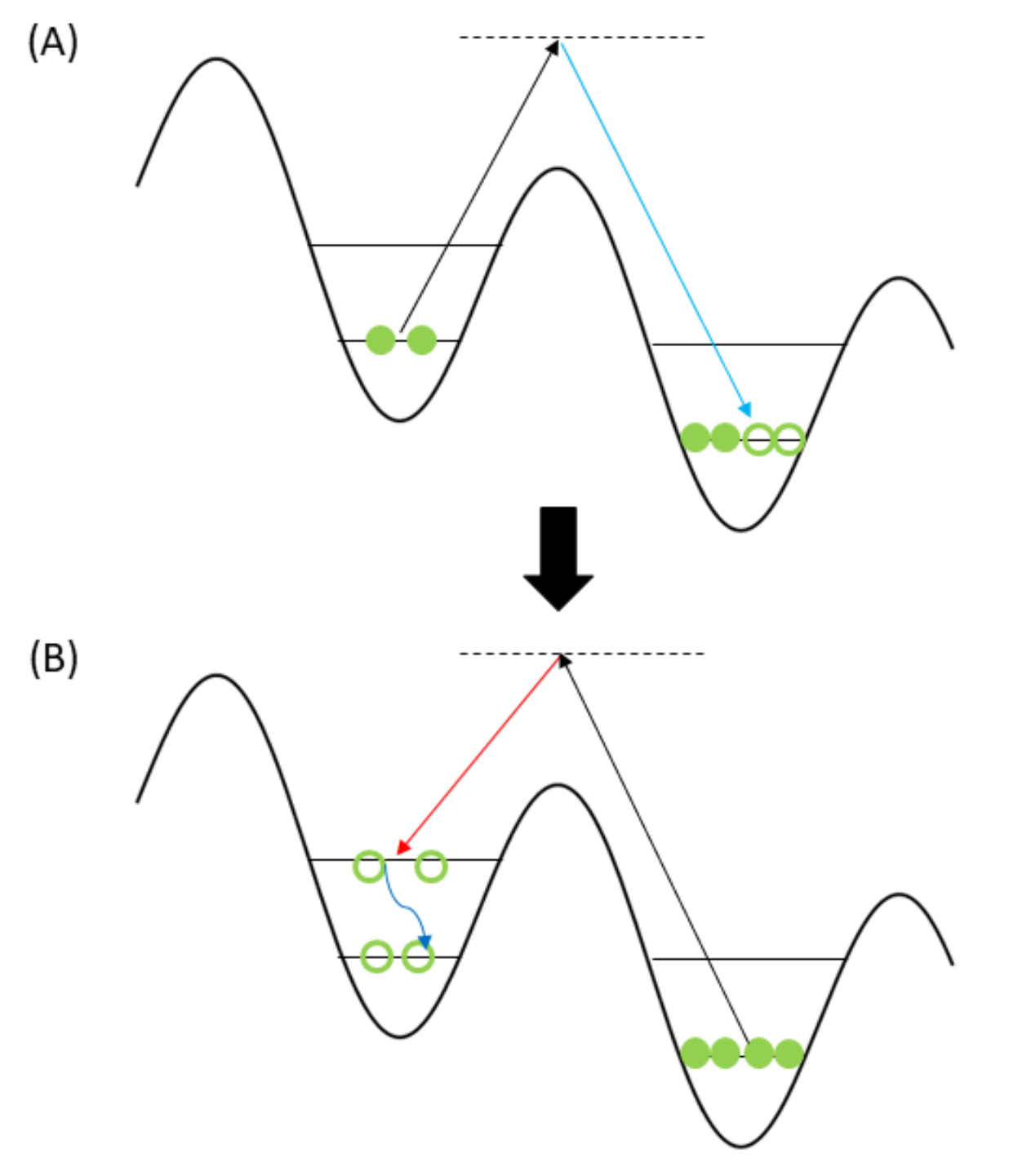}
\caption{
(A)
Coherent resonant Raman transfer of two atoms from the left site to the right site. (B) Coherent resonant Raman transfer of the two atoms back to the excited band on the left site. The two atoms then decay to the ground band.  Due to bosonic symmetry, this sequence is only possible if the total spin of the four atoms is $S_T=2$, and hence this process cannot occur in the AKLT state.}
\label{AKLTfig}
\end{figure}


The protocol is shown in Fig.~\ref{AKLTfig}. 
We switch on a coherent Raman drive such that it is in resonance with transitions where two atoms from the ground band of one site are transported to the ground band of the (filled) site next to it down the ladder, resulting in four atoms on that site. 
Due to Bose statistics, four spin-1/2 bosons on the same site in the same band must be in a total spin symmetric state.  Thus this Raman process can only occur if the atoms on neighboring sites are in an
$S_T = 2$ state.  This process cannot occur if we are in the AKLT state.

The initial, final, and intermediate states in this process have $(2,2),(0,4),$ and $(1,3)$ atoms on the two sites.  They have interaction energies $2U_{gg},6 U_{gg},$ and $2 U_{gg}$.  If the drive is chosen so that the initial and final states are resonant, then the intermediate state is detuned by $\Delta-3 U_{gg}$.  The process therefore occurs at a rate
%
%
%
\begin{equation}\label{fourphoton}
  \Omega^{''}\sim \frac{\Omega^{\prime 2}}{|\Delta-3U_{gg}|}. 
\end{equation}
As in Sec.~\ref{drivesection}, $\Omega^{'}$ is the rate of coherently transferring one atom by one site.

We would operate in the limit where $U_{gg}, \Delta \gg \Omega^{''} \gtrsim \Delta U, \Gamma$. This hierarchy of energy scales ensures that unwanted processes are far off resonant.

Simultaneously, another set of Raman lasers 
resonantly transfers two atoms from the four-atom site back to their original site but in the excited motional band as shown in Fig.~\ref{AKLTfig}. Following our previous arguments, the rate of this process would scale as $\Omega^{''}\sim \Omega^{\prime 2}/|\Delta-3U_{gg}+\omega_0|$. Here the intermediate state with three atoms on one of the sites has an extra detuning $\omega_0$ due to the atoms being transferred into the excited band. The same hierarchy of energy scales makes only this process resonant for all $S_z$ combinations. The two atoms in the higher band can now decay through their interaction with the superfluid bath atoms.

Spin decoherence is engineered through two mechanisms.  First, there is a dephasing resulting from time spent in the excited state configuration.  For concreteness, consider a neighboring pair in the $|S_T=2,S_T^Z=1\rangle$ state, $\frac{1}{\sqrt{2}}(|+_e 0\rangle + |0_e +\rangle)$ with atoms on one of the sites in the excited band.
The $|+_e 0\rangle$ state  has energy $U_{ee}^{|+\rangle}+U_{gg}^{|0\rangle}$,
while the $|0_e +\rangle$ state has energy
$U_{ee}^{|0\rangle}+U_{gg}^{|+\rangle}$. These two states have an energy difference given by, $\Delta U_{gg}-\Delta U_{ee}-(1/4)\Delta U$. Thus their relative phases wind at a rate $(1/4) \Delta U$.  As long as $ \Gamma^{\prime} {\lesssim} (1/4)\Delta U$, then the phase is effectively scrambled.  Here $\Gamma^{\prime}=\Gamma/2$ represents the effective decay rate of the two atoms from the motionally excited band.

A second mechanism for decoherence comes from the decay process itself.  The transition $|+_e\rangle\to|+\rangle$ requires emitting a different frequency Bogoliubov phonon than the transition $|0_e\rangle\to|0\rangle$.  
The  bath learns which-path information, and 
this is then analogous to a measurement.  For the energy difference to be resolvable, we require $\Gamma^{\prime} { \lesssim} (1/4)\Delta U$.




After the decay, the probability of being in the $S_T=2$ sector has been reduced.
These processes would continue until no neighboring pairs are in the $S_T=2$ channel. The chain is then in the AKLT state, and all dynamics stop.

We used a Lindblad master equation approach to model the dynamics. 
As in our previous treatments, we do not explicitly model the various intermediate states, and instead work with an effective model, where the incoherent processes are described by jump operators of the form
%
%
%
%
\begin{equation}\label{jump}
    \hat{C}_{i,kk^\prime} = \sqrt{\Gamma^{\prime}}|k k^\prime\rangle\langle kk^\prime| \hat{P}_{i,i+1}^{S_T=2}
\end{equation}
Here, $k, k^\prime \in \{|+\rangle , |0\rangle , |-\rangle\}$, and $\hat{P}_{i,i+1}^{S_T=2}$ projects the spin-1 objects on neighboring sites to the spin-2 sector. The effective rate is $\Gamma^{\prime}$.

The master equation for the density matrix $\hat{\rho}$ 
is
%
\begin{equation}\label{lindblad}
    \frac{d\hat{\rho}}{dt} = \sum_{i=1}^{i=N-1}\sum_{kk^\prime} \hat{C}_{i,kk^\prime}\hat{\rho}\hat{C}_{i,kk^\prime}^{\dagger } -\frac{1}{2}\{\hat{C}_{i,kk^\prime}^{\dagger}\hat{C}_{i,kk^\prime},\hat{\rho}\}
\end{equation}
By construction, 
superpositions of the four AKLT states are the only steady states.

We use two approaches for analyzing the behavior:  in Sec.~\ref{diag}, we numerically calculate the eigenvalues of the Lindblad super-operator.  The real part of the smallest non-zero eigenvalue gives the time-scale for approaching the AKLT state. The size of the matrix we need to diagonalize grows exponentially with the size of the system, limiting this technique to chains with fewer than 7 sites.

In Sec.~\ref{dmrg}, we instead use a stochastic wavefunction approach which is equivalent to
Eq.~(\ref{lindblad}).  We write the wavefunction as a matrix product state, and use tensor network tools to efficiently evolve it in time.  
We take the initial state as a product state of $|0\rangle$ on all sites.
We measure the expectation value of the sum of all the nearest neighbor spin-2 projectors.  At long times, this decays exponentially -- and we extract the time-scale by fitting this exponential.

We find that the time to create the AKLT state scales as $(N-1)^2$, where $N$ is the number of sites.  We give an intuitive understanding of this result based upon diffusion of domain walls.

%

\subsubsection{Exact Diagonalization}\label{diag}

We vectorize the density matrix by putting all of its elements 
in a column vector, denoted $\tilde \rho$.  
The Lindblad equation then has the structure of a linear differential equation with constant coefficients and we can
use standard linear algebra techniques to find the rate of approaching equilibrium.

If we do not take advantage of any symmetries, our Hilbert space has length $3^N$, where $N$ is the number of spins.  The density matrix is a $3^N \times 3^N$ matrix, so $\tilde\rho$ is a vector of length $3^{2N}$.  The index $\alpha$ which labels the elements of $\tilde \rho$ is associated with a bra $\langle\psi|$ and a ket $|\phi\rangle$, and 
$\tilde \rho_\alpha = \langle\psi|\hat\rho|\phi\rangle_\alpha$. Here $\langle\psi|$ and $|\phi\rangle$ are arbitrary states in the $3^N$ dimensional Hilbert space.

In its vectorized form, the Lindblad equation from Eq.~(\ref{lindblad}) is
\begin{equation}
     \frac{d\tilde\rho_\alpha}{dt} =\sum_{\beta} \mathcal{\hat{L}_{\alpha\beta}}\tilde\rho_\beta
\end{equation}
where the 
the matrix on the right has elements
\begin{equation}
\mathcal{\hat{L}}_{\alpha\beta}=\big(| \psi\rangle\otimes\langle\phi|\big)_\alpha \mathcal{\hat{L}}\big(
|\psi\rangle\otimes\langle\phi|\big)_\beta.
\end{equation}
The
Lindblad
superoperator $\mathcal{\hat{L}}$ is
\begin{equation}
\begin{split}
    \mathcal{\hat{L}}= \sum_{i=1}^{i=N-1}\sum_{kk^\prime} \hat{C}_{i,kk^\prime}\otimes \hat{C}_{i,kk^\prime} -\frac{1}{2}\hat{C}_{i,kk^\prime}^{\dagger}\hat{C}_{i,kk^\prime}\otimes \mathds{1}\\
    -\frac{1}{2}\mathds{1}\otimes \hat{C}_{i,kk^\prime}^{\dagger}\hat{C}_{i,kk^\prime}.  
\end{split}
\end{equation}



The non-zero eigenvalues of 
$\mathcal{\hat{L}}$ have negative real parts which give
the rates of decay of various perturbations.  The zero eigenvalues identify the dark states.
The total $S_z$ of the chain is conserved in the dynamics, 
so $\mathcal{\hat{L}}$ is block diagonal.
We restrict ourselves to the block with $S_z=0$.  


We find four zero-eigenvalues, corresponding to two of the AKLT states, and the coherences between them.  These AKLT states have edge modes 
$|\uparrow \downarrow \rangle$ and $|\downarrow \uparrow \rangle$.

The time taken to reach the AKLT state is controlled by the eigenvalue whose real part has the smallest non-zero magnitude, $\gamma$.
Figure~\ref{times} shows how this slowest rate scales with $N$ for the exact diagonalization calculation.  Due to the exponential scaling of the Hilbert space, we are restricted to $N< 7$.

As will be discussed in more detail in Sec.~\ref{dmrg}, the rate scales inversely with the number of sites as $\gamma\propto 1/(N-1)^2$. 

\subsubsection{DMRG Calculation}\label{dmrg}

To explore larger systems,  we use the stochastic wavefunction formalism \cite{openq}. In this approach, one uses a non-Hermitian effective Hamiltonian to evolve an initial wavefunction in time, stochastically including discrete ``quantum jumps.'' The non-Hermitian part of the Hamiltonian is constructed using the jump operators and leads to a non-unitary evolution of the wavefunction. The stochastic quantum jumps represent the effects of random `measurements' by the environment through the application of a random jump operator onto the wavefunction. 

Following the approach introduced by Daley et al.~\cite{mpstrajec1}, and also used by Bonnes and L\"auchli \cite{mpstrajec2}, we use a matrix product state ansatz for the time dependent wavefunction.  This is extremely efficient, allowing us to model systems with as many as 9 sites (corresponding to a Hilbert space with almost 20,000 states,  a density matrix with almost 400 million elements, and a super-operator with over $10^{17}$ elements). 

In particular, we take
\begin{equation}
|\psi(t)\rangle = \sum_{\{i_j,\sigma_j\}}
A^{(1)}_{1,i_1,\sigma_1} A^{(2)}_{i_1, i_2,\sigma_2}\cdots A^{(n)}_{i_{n-1},1,\sigma_n}\,|\sigma_1\cdots\sigma_n\rangle.
\end{equation}
The bond indices $i_j$ are dummy variables  which take on no more than $\chi$ different values, where $\chi$ is referred to as the {\em bond dimension}.  The indices   $\sigma_j=-1,0,1$ corresponds to the physical spins.  All time dependence is contained in the tensors $A^{(j)}_{i_{j-1},i_j,\sigma_j}$.  The AKLT state can be represented in this form with $\chi=2$.

We take an initial product state of $|0\rangle$ on all sites. We discretize time.  During each time-step, we evolve the wavefunction as,
$|\psi(t+\delta t)\rangle = (\mathds{1}-i\hat H_{\rm eff}\delta t)|\psi(t)\rangle$, where $\hat H_{\rm eff}= \hat H_0-i\sum_{i,kk^\prime} \hat C^\dagger_{i,kk^\prime} \hat C_{i,kk^\prime}$ is a non-Hermitian effective Hamiltonian.  In our model $\hat H_0=0$ as all dynamics simply comes from the jump operators.

We construct the time evolution operator
 $\hat{O}=(\mathds{1}-i \hat H_{\rm eff} \delta t)$ as a matrix product operator,
\begin{equation}
\begin{split}
\hat{O} = \sum_{\{k_j,\sigma^{\prime}_j,\sigma_j\}}W^{(1)}_{1,k_1,\sigma^{\prime}_1\sigma_1} W^{(2)}_{k_1,k_2,\sigma^{\prime}_2\sigma_2}\cdots\\
\cdots W^{(n)}_{k_{n-1},1,\sigma^{\prime}_n\sigma_n}\,|\sigma_1^{\prime}\cdots\sigma_n^{\prime}\rangle \langle \sigma_1\cdots\sigma_n|.
\end{split}
\end{equation}
Here again the $k_j$ are bond indices and $\sigma_j,\sigma^{\prime}_j$ are physical spins. Our effective Hamiltonian only has nearest-neighbor terms, and the tensors $W^{(i)}$ take the standard form, where blocks of non-zero elements appear on the first column and last row  \cite{dmrgmps}.  To evolve the wavefunction $|\psi(t)\rangle$ with $\hat{O}$, we take tensor products of $A^{(i)}$ with $W^{(i)}$. 
The bond dimension of the evolved wavefunction increases after each step. We use the zip-up method described in~\cite{dmrg} to control the bond-dimension of the resulting state at each time step and proceed with the time evolution.

Due to the non-unitary evolution of the wavefunction by $\hat H_{\rm eff}$, the wavefunction norm is reduced. We calculate this norm, $1-p=\langle \psi(t+\delta t|\psi(t+\delta t)\rangle$.  We then draw a random number $x$ between 0 and 1. If $x>p$, we normalize $|\psi\rangle$, then continue with the next time step.  

If $x<p$ it means a quantum jump has occurred.  We then calculate the probabilities, 
$p_{i,kk^{\prime}} = \langle\psi(t)|\hat C^\dagger_{i,kk^\prime} \hat C_{i,kk^\prime}|\psi(t)\rangle$ where $p=\sum_{i,kk^{\prime}}p_{i,kk^{\prime}}$, and draw another random number to determine which has occurred \cite{openq}.  We apply the relevant jump operator and renormalize the state.



\begin{figure}
\includegraphics[width=\columnwidth]{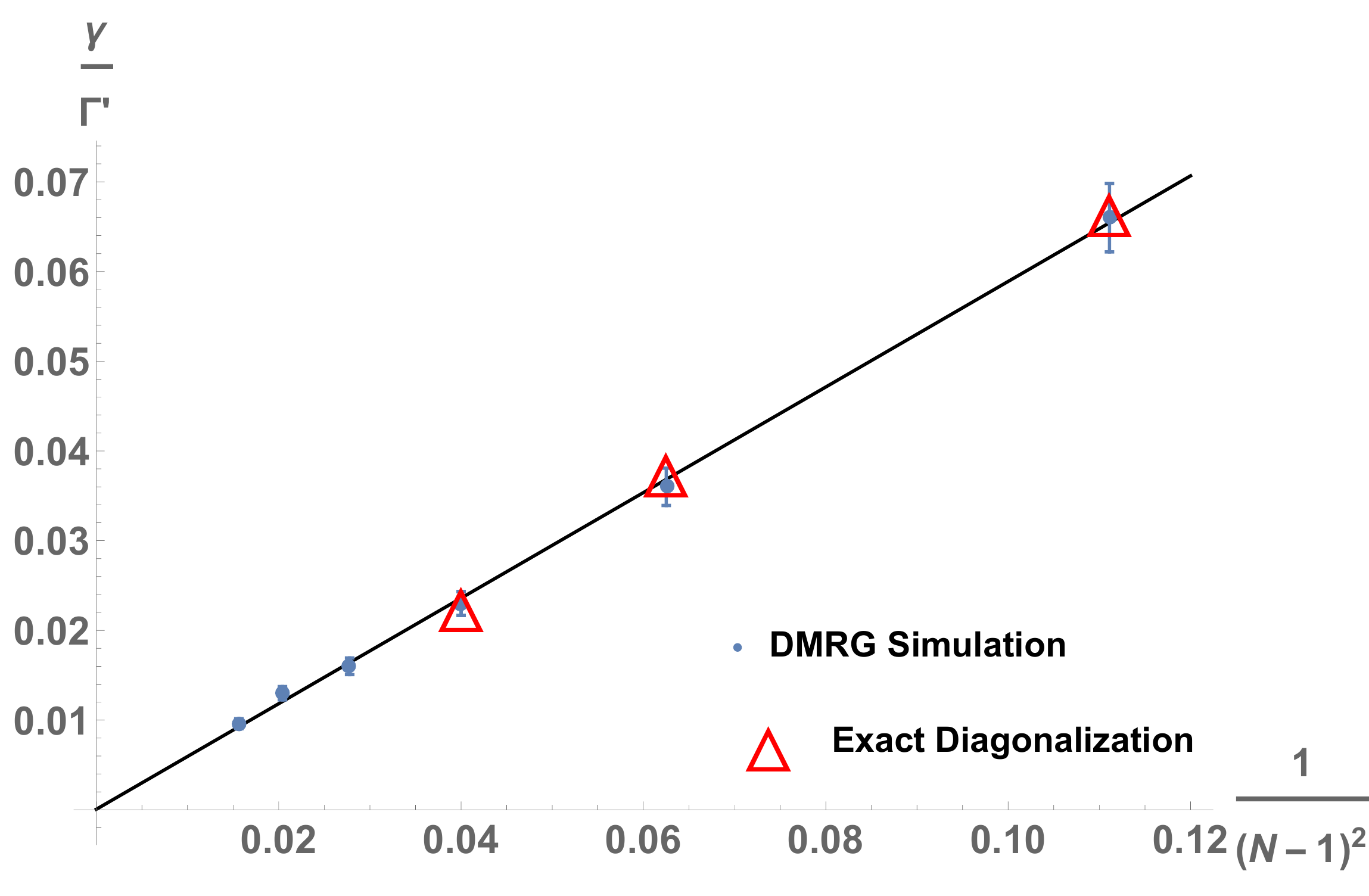}
\caption{(color online) Scaling of slowest rate of decay, $\gamma$, with the number of spins, $N$.  Large red triangles:  Diagonalization of the Lindblad super-operator; Blue dots with error bars:  Stochastic DMRG simulation.  The error bars represent the statistical uncertainty from using 75 realizations in  the DMRG  calculation. The decay is measured in terms of
$\Gamma^{\prime}$, the  rate of the local dissipation.
}
\label{times}
\end{figure}

We measured the total spin-2 projection of all nearest neighbor pairs as a function of time and fit the tail with a decaying exponential function.  Our resulting estimate for the slowest decay rate, $\gamma$ is plotted in Fig.~\ref{times} for up to 9 sites. The DMRG simulation reproduces the exact diagonalization rates and shows the same $1/(N-1)^2$ scaling.   We use 75 realizations, and error bars in Fig.~\ref{times} correspond to the statistical uncertainty in $\gamma$. 


This $1/(N-1)^2$ scaling can be qualitatively understood by analyzing how string order develops in the spin chain.  The AKLT state in the spin-1 basis is a superposition of different arrangements of $|+\rangle$, $|0\rangle$ and $|-\rangle$ -- for example, with three sites, an AKLT state is $\frac{1}{2}|000\rangle - |+-0\rangle -|0+-\rangle +|+0-\rangle$.  There is a "string order" here in that if you threw away all of the spin-0 sites, each of these terms correspond to an antiferromagentic arrangement $|+-\rangle$.  This same property occurs for longer chains.


Domain walls in the string order can be assigned a ``charge" corresponding to the excess local magnetization:  a configuration $|-++-\rangle$ has a positively charged domain wall, and  $|+--+\rangle$ has a negatively charged domain wall.
Our stochastic process involves local projections, which conserve the total magnetization, and hence
cannot remove an isolated 
domain wall. Instead, during the stochastic process, the domain walls undergo random walks -- and can be annihilated when two oppositely charged domain walls touch.  For example, to establish string order in the state $|+\circled{
+
}00
\circled{$-$}
-\rangle$, the circled spins must either exchange position, or annihilate each-other.

The underlying domain wall dynamics are diffusive, and the slowest processes involve the motion of a domain wall over a distance of order $N-1$, where $N$ is the number of sites.  Thus one expects the time required to scale as $(N-1)^2$, as seen in the numerics.

We conclude that the 
state preparation time would scale quadratically with the number of sites. This scaling  is quite favorable.  In contrast, for an adiabatic preparation scheme, one expects the smallest gap to scale exponentially in the system size, and hence the preparation time would also scale in that manner.

\section{Summary and Outlook}

We present concrete examples which elucidate how engineered dissipation, in conjunction with an appropriate drive,  can be used to manipulate bosonic atoms  in tilted optical lattices. In these examples, the drive is supplied by coherent Raman lasers, and the  dissipation comes from band-changing  collisions with a superfluid bath. Using these ingredients, we present 
protocols 
for controlling transport, 
forming a Mott insulator state, and creating the topologically ordered spin-1 AKLT state. In the latter two cases, the states are autonomously stabilized, and any perturbations can be automatically healed.  

We calculate the relevant time scales for state preparation and their dependence on system size. We note that in all cases, the preparation time scales polynomially with system size.  By contrast, adiabatic state preparation techniques typically scale exponentially.

All of the ingredients in our protocols have been individually realized in existing experiments.  Moreover, the three examples form a natural progression for an experimental program:  each adding a layer of sophistication to the previous one.  While all three examples are important, the observation of the AKLT state would be particularly impactful.  

As previously stated, these examples are also important for the way they exemplify general principles of the driven-dissipative manipulation of quantum states.  In describing them, we are able to address the interplay of different energy scales, and rates of processes. We show how the symmetries of the system and the target quantum state can be exploited for dissipative state preparation. The techniques are readily extended into other systems, and into other forms of manipulation.

While all of the ingredients have been realized in experiments,
a real challenge with our proposals is that they require combining a number of sophisticated experimental techniques.  The superfluid bath, which  provides our dissipation, is one of the most challenging elements. 
It would be desirable to construct an all-optical scheme which uses spontaneous photon emission to provide the dissipation.  As discussed in Sec.~\ref{limitations}, the difficulty there is that optical line-widths are larger than the other scales in the system, and would lead to unwanted transitions. 
An important future 
research
direction would be to explore dressed state techniques or other ways to
overcome this difficulty. 
\section*{Acknowledgements}  This work was supported by the NSF Grant PHY-1806357.  EJM would like to acknowledge discussions with Wolfgang Ketterle and Andrew Daley during this research's formative stages.
\\
\appendix

\section{Raman Rates}\label{ramanapp}
Here we derive the expression in Eq.~(\ref{ramanratefinal}) for the Raman matrix elements.  We begin with Eq.~(\ref{ramanrate}),
\begin{equation}
\Omega^{\prime} = \frac{\Omega_0^2}{\delta}
    \left|\int dx\,
    \psi_{m,g}^*(x)
    e^{i(k_1-k_2)x}\psi_{m+1,e}(x) 
    \right|^2
\end{equation}

In the limit where $\Delta \gg t_{\alpha}$, we can approximate the Wannier-Stark eigenstates in Eq.~(\ref{WSfns}) upto leading order as,
\begin{equation}
    \psi_{m,\alpha}(x) \sim \phi_m^\alpha(x)
\end{equation}

As described in Sec.~\ref{1dlat}, in the deep lattice limit, $\phi_m^\alpha(x)\approx
w_\alpha(x-m d/2)$, where $w_\alpha(x-m d/2)$ are highly localized harmonic oscillator eigenstates. The Raman matrix element becomes, 
\begin{equation}
    \Omega^\prime = \frac{\Omega_0^2}{\delta}|I|^2
\end{equation}
where, 
\begin{equation}
I = \int dx\, w_{g}^*(x-md/2)e^{ikx}w_{e}(x-(m+1)d/2).
\end{equation}
Here $k=k_1 - k_2$ where $k_1$ and $k_2$ are the wavevectors of the two raman lasers. The Gaussian integral is readily calculated,
\begin{equation}
   I = \left(iC -\frac{\pi V_0^{1/4}}{\sqrt{2}}\right) e^{-\frac{C^2}{2}-\frac{ikd}{4} -\frac{\pi^2 \sqrt{V_0}}{4}}
\end{equation}
with
$C={k d}/\left({2\sqrt{2} \pi V_0^{1/4}}\right)$.

For a sufficiently deep lattice, the net transition rate scales as,
\begin{equation}
    \Omega^\prime \sim  \frac{\Omega_0^2}{\delta}\left(\frac{k^2d^2}{8\pi^2\sqrt{V_0}}+ \frac{\pi^2\sqrt{V_0}}{2} \right)  e^{-\frac{k^2d^2}{8\pi^2 \sqrt{V_0}}-\frac{\pi^2\sqrt{V_0}}{2}}
\end{equation}

Many of the coefficients are under experimental control. $\Omega_0$ is the dipole matrix element between the ground and excited electronic levels of the lattice atoms, which directly depends on the laser intensity. The wavevector $k$ is tuned by changing the angle between the two lattice beams, the optimal value is given by $k^2=8\pi^2 \sqrt{V_0}/d^2$, in which case
\begin{equation}
\Omega^\prime \sim
\frac{\Omega_0^2}{\delta} \left(1+ \frac{\pi^2\sqrt{V_0}}{2} \right) e^{-\frac{\pi^2\sqrt{V_0}}{2}}. 
\end{equation}


\begin{thebibliography}{30}
\bibitem{coolmethod} D. McKay and B. DeMarco, \textit{Cooling in strongly correlated optical lattices: prospects and challenges}, Rep. Prog. Phys. 74, 054401 (2011)
\bibitem{engnrdissipation} M. M¨uller, S. Diehl, G. Pupillo and P. Zoller, \textit{Engineered Open Systems and Quantum Simulations with Atoms and Ions}, Advances In Atomic, Molecular, and Optical Physics, 61 (2012), pp. 1–80
\bibitem{adiabatictheorem} M. H. S. Amin, \textit{Consistency of the Adiabatic Theorem}
Phys. Rev. Lett. 102, 220401 (2009)
\bibitem{mottphoton} Ruichao Ma, Brendan Saxberg, Clai Owens, Nelson Leung, Yao Lu, Jonathan Simon and David I. Schuster, \textit{A dissipatively stabilized Mott insulator of photons}, Nature 566, 51–57 (2019)
\bibitem{bell} S. Shankar, M. Hatridge, Z. Leghtas, K. M. Sliwa, A. Narla, U. Vool, S. M. Girvin, L. Frunzio, M. Mirrahimi and M. H. Devoret, \textit{Autonomously stabilized entanglement between two superconducting quantum bits}, Nature 504, 419–422 (2013).
\bibitem{GHZ} Julio T. Barreiro, Markus Müller, Philipp Schindler, Daniel Nigg, Thomas Monz, Michael Chwalla, Markus Hennrich, Christian F. Roos, Peter Zoller and Rainer Blatt, \textit{An open-system quantum simulator with trapped ions}, Nature 470, 486–491 (2011).
\bibitem{bosehubbard} T. Tomita, S. Nakajima, I. Danshita, Y. Takasu and Y. Takahashi, \textit{Observation of the Mott insulator to superfluid crossover of a driven-dissipative Bose-Hubbard system}, Sci. Adv. 3, e1701513 (2017).
\bibitem{squeeze} Roland Cristopher F. Caballar, Sebastian Diehl, Harri Mäkelä, Markus Oberthaler and Gentaro Watanabe, \textit{Dissipative preparation of phase- and number-squeezed states with ultracold atoms}, Phys. Rev. A 89, 013620 (2014)
\bibitem{toposupercond} Fernando Iemini, Davide Rossini, Rosario Fazio, Sebastian Diehl and Leonardo Mazza, \textit{Dissipative topological superconductors in number-conserving systems}, Phys. Rev. B 93, 115113 (2016)


\bibitem{tilt} Ivana Dimitrova, Niklas Jepsen, Anton Buyskikh, Araceli Venegas-Gomez, Jesse Amato-Grill, Andrew Daley and Wolfgang Ketterle, \textit{Enhanced Superexchange in a Tilted Mott Insulator}, Phys. Rev. Lett. 124, 043204, (2020)
\bibitem{maggradient} J. Simon, W. S. Bakr, R. Ma, M. E. Tai, P. M. Preiss and M.
Greiner, \textit{Quantum simulation of antiferromagnetic spin chains in an optical lattice}, Nature (London) 472, 307 (2011)
\bibitem{WScoherent} Q. Beaufils, G. Tackmann, X. Wang, B. Pelle, S. Pelisson, P. Wolf and F. Pereira dos Santos, \textit{Laser Controlled Tunneling in a Vertical Optical Lattice}, Phys. Rev. Lett. 106, 213002 (2011)
\bibitem{NaLi} R. Scelle, T. Rentrop, A. Trautmann, T. Schuster and M. K. Oberthaler, \textit{Motional Coherence of Fermions Immersed in a Bose Gas}, Phys. Rev. Lett. 111, 070401 (2013)
\bibitem{DeMarcobath} D. Chen, C. Meldgin and B. DeMarco, \textit{Bath-induced band decay of a Hubbard lattice gas}, Phys. Rev. A 90, 013602 (2014)

\bibitem{Zollerbath} A Griessner, A J Daley, S R Clark, D Jaksch and P Zoller,\textit{Dissipative dynamics of atomic Hubbard models coupled to a phonon bath: dark state cooling of atoms within a Bloch band of an optical lattice}, New J. Phys. Volume 9, 44 (2007)

\bibitem{specdeplatt} L. J. LeBlanc and J. H. Thywissen, \textit{Species-specific optical lattices}, Phys. Rev. A 75, 053612 (2007)

\bibitem{Wstark} David Emin and C.F. Hart, \textit{Existence of Wannier-Stark localization}, Phys. Rev. B 36, 7353, (1987).
\bibitem{Liswave} E. R. I. Abraham, W. I. McAlexander, C. A. Sackett and Randall G. Hulet, \textit{Spectroscopic Determination of the s-Wave Scattering Length of Lithium}, Phys. Rev. Lett. 74, 1315 (1995)



\bibitem{wall} Alexander L. Gaunt, Tobias F. Schmidutz, Igor Gotlibovych, Robert P. Smith and Zoran Hadzibabic, \textit{Bose-Einstein Condensation of Atoms in a Uniform Potential}, Phys. Rev. Lett. 110, 200406 (2013)
\bibitem{mottstate} D. van Oosten, P. van der Straten and H. T. C. Stoof, \textit{Quantum phases in an optical lattice},
Phys. Rev. A, 63, 053601 (2001).
\bibitem{gillespie} Gillespie, Daniel T. (1976). "A General Method for Numerically Simulating the Stochastic Time Evolution of Coupled Chemical Reactions". Journal of Computational Physics. 22 (4): 403–434.
\bibitem{doublon} F. Meinert, M.J. Mark, K. Lauber, A.J. Daley and H.-C. Nägerl, \textit{Floquet Engineering of Correlated Tunneling in the Bose-Hubbard Model with Ultracold Atoms}, Phys. Rev. Lett. 116, 205301 (2016)
\bibitem{Ketterletwocomp} Jesse Amato-Grill, Niklas Jepsen, Ivana Dimitrova, William Lunden, and Wolfgang Ketterle, \textit{Interaction spectroscopy of a two-component Mott insulator}, Phys. Rev. A 99, 033612 (2019)
\bibitem{aklt} Ian Affleck, Tom Kennedy, Elliott H. Lieb and Hal Tasaki, \textit{Rigorous results on valence-bond ground states in antiferromagnets}, Phys. Rev. Lett. 59, 799 (1987)
\bibitem{quantumcomputing}Gavin K. Brennen and Akimasa Miyake, \textit{Measurement-Based Quantum Computer in the Gapped Ground State of a Two-Body Hamiltonian}, Phys. Rev. Lett. 101, 010502 (2008)
\bibitem{stabilizer} Daniel Gottesman, \textit{Stabilizer Codes and Quantum Error Correction}, arXiv:quant-ph/9705052
\bibitem{openq} Andrew J. Daley, \textit{Quantum trajectories and open many-body quantum systems} , Advances in Physics, 63:2, 77-149 (2014)
\bibitem{mpstrajec1} A. J. Daley, J. M. Taylor, S. Diehl, M. Baranov, and P. Zoller, \textit{Atomic Three-Body Loss as a Dynamical Three-Body Interaction}, Phys. Rev. Lett. 102, 040402 (2009)
\bibitem{mpstrajec2} Lars Bonnes and Andreas M. Läuchli, \textit{Superoperators vs. Trajectories for Matrix Product State Simulations of Open Quantum System: A Case Study}, arXiv:1411.4831 (2014)

\bibitem{dmrgmps}
	U. Schollwöck, \textit{The density-matrix renormalization group in
the age of matrix product states}, Ann. Phys. (NY) 326, 96
(2011).

\bibitem{dmrg} S. Paeckel, T. Köhler, A. Swoboda, S. R. Manmana, U.
Schollwöck, and C. Hubig, \textit{Time-evolution methods for matrix product
states}, Ann. Phys. (NY) 411, 167998 (2019).



\end{thebibliography}
\end{document}